\documentclass[useAMS,usenatbib]{mn2e}
\usepackage{graphics,epsf,graphicx}

\def \OIII {[O{\sc iii}]~5007 \AA}
\def \vhel{\ifmmode{~v_{{\rm HEL}}}\else{~$v_{{\rm HEL}}$}\fi}
\def \vsys{\ifmmode{~v_{{\rm sys}}}\else{~$v_{{\rm sys}}$}\fi}
\def \HA {\ifmmode{{\rm H}\alpha~}\else{${\rm H}\alpha$~}\fi}
\def \kms{\ifmmode{~{\rm km\,s}^{-1}}\else{~km~s$^{-1}$}\fi}
\def \msun{\ifmmode{{\rm M}_\odot}\else{${\rm M}_\odot$}\fi}

\title[Evidence for ablated flows in the shell of nova DQ Her]{Evidence for ablated flows in the shell of nova DQ Her}

\author[N. M. H. Vaytet, T. J. O'Brien and A. P. Rushton]{N. M. H. Vaytet\footnotemark[1], T. J. O'Brien and A. P. Rushton \\Jodrell Bank Observatory, School of Physics and Astronomy, The University of Manchester, Macclesfield, Cheshire, SK11 9DL, UK}

\begin{document}

\date{To appear in the Monthly Notices of the Royal Astronomical Society}

\pagerange{\pageref{firstpage}--\pageref{lastpage}} \pubyear{2007}

\maketitle

\label{firstpage}

\begin{abstract}
High-resolution longslit \HA spectra of the shell of the old nova DQ Her have been obtained with the William Herschel Telescope using the ISIS spectrograph. An equatorial expansion velocity of $370\pm14$\kms ~is derived from the spectra which, in conjunction with a narrowband \HA image of the remnant, allows a distance estimate of $525\pm28$pc. An equatorial ring which exhibits enhanced [N{\sc ii}] emission has also been detected and the inclination angle of the shell is found to be $86.8 \pm 0.2$ degrees with respect to the line of sight. The spectra also reveal tails extending from the clumps in the shell, which have a radial velocity increasing along their length. This suggests the presence of a stellar wind, collimated in the polar direction, which ablates fragments of material from the clumps and accelerates them into its stream up to a terminal velocity of order 800--900\kms.
\end{abstract}


\footnotetext[1]{email: neil.m.vaytet@postgrad.manchester.ac.uk}

\section{Introduction}\label{s1}

DQ Herculis is a classical nova that outburst in 1934. It was first detected on December 12 of the same year at a magnitude of 3.3 \citep{campbell35}. It then reached a maximum apparent magnitude of 1.4 and declined with a t$_{3}$ of 94 days (t$_{3}$ is the time it takes the nova to decline 3 magnitudes in brightness from peak) (\citealt{gaposchkin56}; see also \citealt{beer74}). DQ Her is the prototype of its own cataclysmic variable sub-class of intermediate polars (or DQ Her stars), distinguished by a rapidly rotating highly magnetic white dwarf (WD) and a companion red star. The powerful magnetic field of the accreting WD (few$\times 10^{6}$ Gauss) disrupts the inner accretion disc, only allowing material to accrete onto its magnetic poles \citep{patterson94}. As matter piles up on the surface of the primary, the increase in pressure and temperature in the degenerate layer of accreted hydrogen eventually leads to a thermonuclear runaway (TNR) on the surface of the WD, resulting in the ejection of a shell of material into the circumstellar medium \citep{starrfield89}.

The components of the central binary system of DQ Her have masses of $M_{WD}=0.6\msun$ and $M_{c}=0.4\msun$ for the WD and the companion respectively \citep{horne93}. The distance to DQ Her is still a matter of debate. Until recently, it was believed to be of order $400\pm100$ pc \citep{young81} and $485\pm50$ pc \citep{martin89}. \citet{herbig92} however, propose a value of $561\pm20$ pc.

In this paper, we present the results from high-resolution longslit spectroscopy of the shell of DQ Her. Expansion velocities, physical size and distance were derived. Recent studies of winds from WDs in cataclysmic variables have also suggested that the accretion disc of the central binary has an important role in the polar collimation of the wind (\citealt{shlosman93}; \citealt{knigge96}). This paper addresses the kinematics of polar tails flowing radially out of the shell. Their origin is believed to be the result of either ablation of clumps situated on the surface of the nova shell by a fast stellar wind, or shadowing from the central photoionising radiation by the dense clumps \citep{slavin95}. We here disambiguate the mechanism by showing that their velocity profiles are accordant to wind ablation.

\section{Observations and results}\label{s2}

\begin{figure*}
\begin{center}
\includegraphics[scale=0.66]{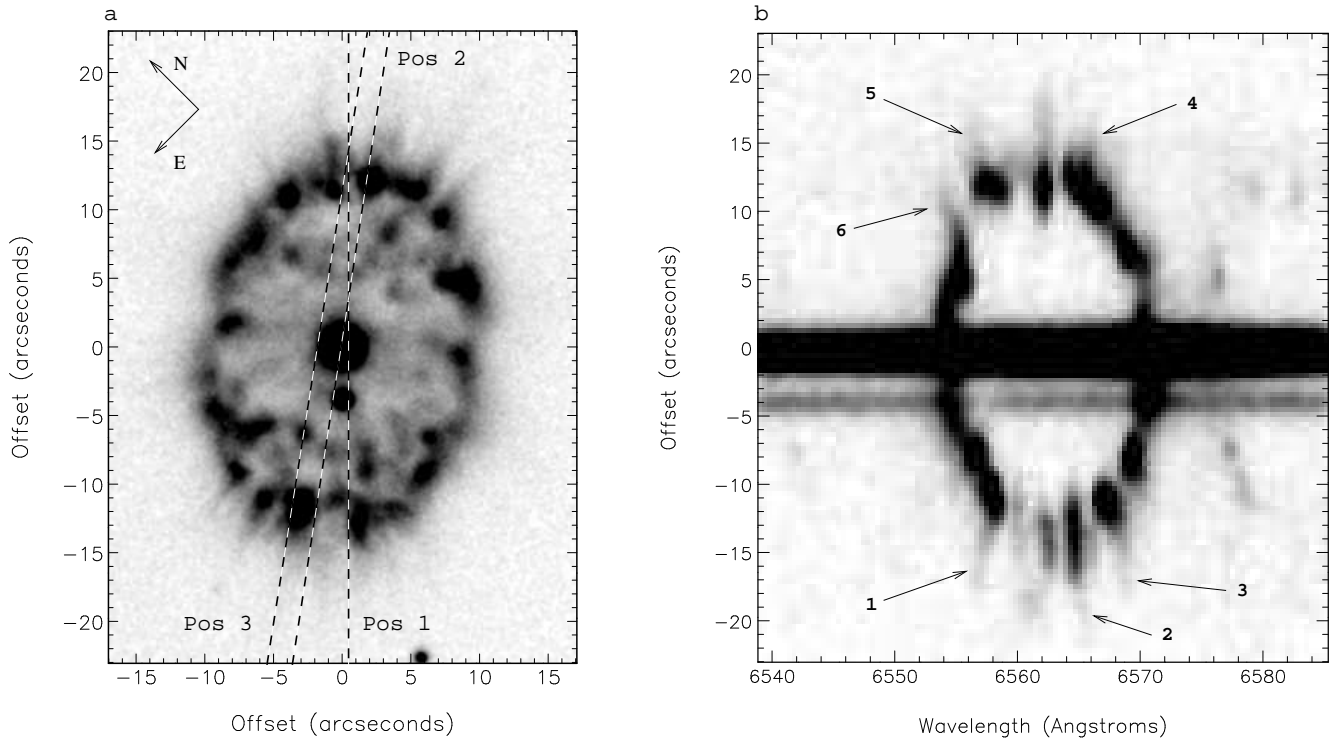}
\end{center}
\caption{(a) The nova remnant DQ Her. The axes show distances in arcseconds relative to the centre of DQ Her. The position of 1" width slits are superimposed on the image (see text for details).\newline
(b) Red-arm ISIS spectrum for slit position 1. The clumpiness in the elliptical shell is clearly visible. We also note the faint presence of the [N{\sc ii}] velocity ellipse centred around $\lambda = 6583.5$\AA ~on the right hand side of the bright \HA emission. The horizontal dark line going through the centre of the image is the continuum emission from the bright central stellar system. The six most visible tails are indicated by the arrows.}
\label{1}
\end{figure*}

Longslit spectra of DQ Her were taken with the ISIS spectrometer on the 4.2 m William Herschel Telescope (WHT) in La Palma, Canary Islands on 1996 August 1--3. This instrument possesses a blue and red arm, allowing simultaneous high-resolution spectroscopy at two different wavelength ranges. The TEK-1 and TEK-2 CCD detectors (both devices have a useful imaging area of $1024\times1024$ pixels for a pixel size of $24\times24 \mu$m) were used with the R1200B and R1200R gratings, achieving a spectral range of approximately 420 \AA ~with a dispersion of 0.41 \AA ~per pixel and central wavelengths of 4950 \AA ~and 6600 \AA ~for the blue and red arms respectively. This permits study of the H$\mathrm{\beta}$ ($\lambda = 4861.3$\AA) and \OIII ~($\lambda = 5006.9$\AA) emission lines at the blue end, and the \HA ($\lambda = 6562.8$\AA) and the two [N{\sc ii}] lines at $\lambda = 6548.1$\AA~ and $\lambda = 6583.5$\AA ~at the red end of the spectrum. The spatial resolution was 0.36 arcseconds/pixel and each spectrum was taken with an exposure time of 1800s. The spectral data were processed using \textsc{starlink} software packages. They were bias and sky subtracted, cleaned from cosmic rays and then wavelength calibrated using a Thorium-Argon calibration lamp.

A 1200s \HA narrowband (filter FWHM = 55\AA) image of the nova was taken at the WHT on October 25 1997 using the TEK-5 CCD chip ($1024\times1024$ $24\times24 \mu$m pixels) and is shown on Fig.~\ref{1} (the image was also de-biased, sky-subtracted and cosmic ray cleaned). The scale of the chip at the f ratio of the camera combined with the WHT was 0.11" per pixel, theoretical resolution which is swamped by seeing effects of the order of $0.6" - 0.7"$ estimated from the FWHM of stellar profiles on the image. It revealed a very irregular, clumpy elliptical shell structure, which had a pinched appearance at the equator, confirming the observations of \citet{herbig92}.

DQ Her is also known to be an eclipsing system \citep{walker56}, which limits the tilt angle of the shell with respect to the line of sight to an inclination $i \geq 80^{\circ}$, assuming the inclination of the shell is the same as the inclination of the binary system. It is in fact believed to be very close to $90^{\circ}$ (\citealt{wood00}; \citealt{herbig92}).

\subsection{General structure}\label{s3}

\begin{figure*}
\begin{center}
\includegraphics[scale=1]{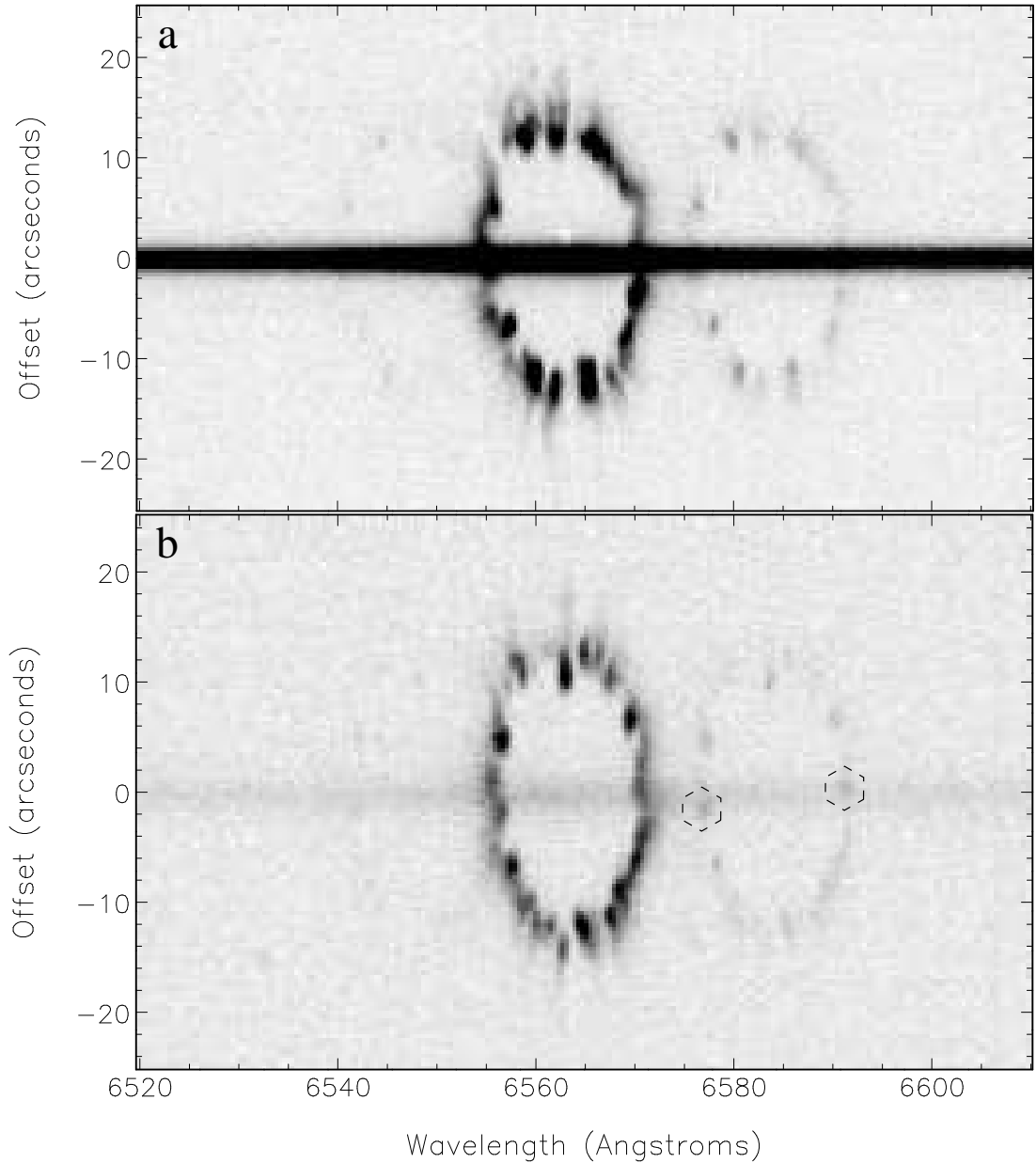}
\end{center}
\caption{(a) \HA + [N{\sc ii}] spectrum for slit position 2. (b) \HA + [N{\sc ii}] spectrum for slit position 3 offset from the central star. The equatorial ring enhancement in [N{\sc ii}] emission is circled (see text).}
\label{2}
\end{figure*}

The slit positions are shown on Fig.~\ref{1}a and the \HA + [N{\sc ii}] spectrum for position 1 is shown in greyscale on Fig.~\ref{1}b. The red-arm spectra for positions 2 and 3 are displayed on Fig~\ref{2}.

\begin{figure}
\begin{center}
\includegraphics[scale=0.47]{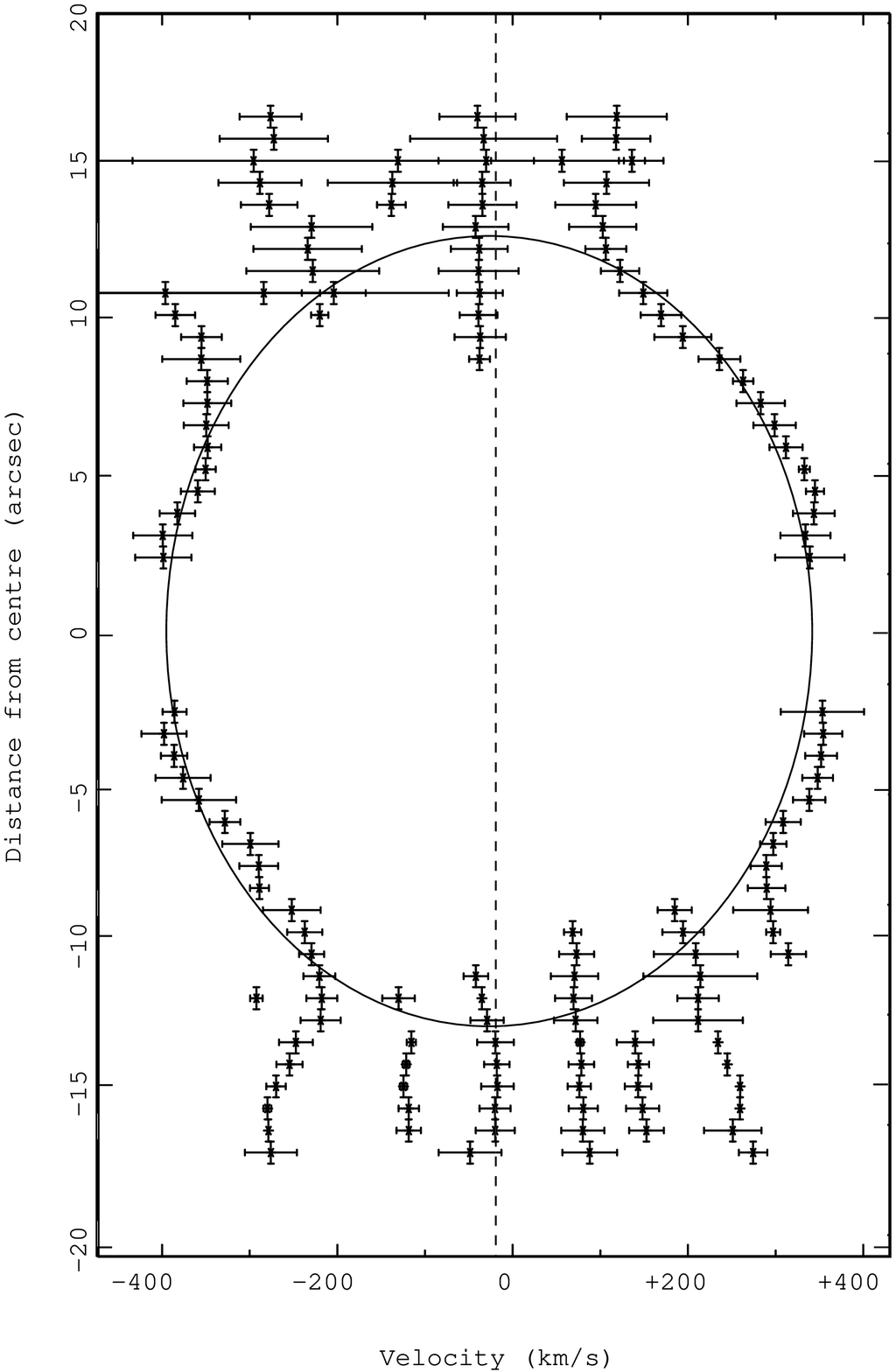}
\end{center}
\caption{\HA velocity profile of DQ Her for the position 1 spectrum. The ellipse shows the p-v array expected for an ellipsoid with the same minor and major axes expanding at a constant rate, with expansion velocity proportional to distance from centre (i.e. Hubble type expansion law). The dotted line going through the centre of the ellipse indicates the systemic velocity of the system at $-20$\kms. Extended tails at the poles are clearly visible. Finally, we can also see emission interior to the shell; see section \ref{s4} for details.}
\label{3}
\end{figure}

Gaussian profiles were fitted to the emission lines using the manual fitting program \textsc{longslit}. The centroids of these Gaussians were then used to calculate the line of sight heliocentric velocity ($v_{hel}$) as a function of position along the slit. The position-velocity (p-v) array for the position 1 \HA emission is displayed in Fig.~\ref{3}.

The image and p-v array are consistent with an expanding ellipsoidal shell seen almost side-on. The ellipse appears to be centred on a systemic heliocentric velocity $\vsys = -20\pm10$\kms. The resulting maximum line of sight velocities were found around the equator with $v_{hel}=\pm(370\pm14)$\kms ~which is close to the expansion velocity of the system since the shell is thought to be viewed at an inclination $i \approx 90^{\circ}$. The expansion velocity of the equatorial belt must be somewhat smaller as the nebula presents a pinched equatorial waist.

The major and minor axes are $a = 25.31\pm0.44$ arcsec and $b=18.70\pm0.44$ arcsec respectively (ignoring the pinch at the equator, and this will be the case for all calculations). The eccentricity of the ellipse is therefore $e = 0.674\pm0.024$.

Assuming a constant expansion rate since the explosion, a distance estimate can be made. The time between outburst in December 1934 and the epoch of the image is approximately $t=62.88\pm0.05$ years. Hence the equatorial radius of the object is $(7.34\pm0.20) \times 10^{14}$ m. Assuming a constant expansion velocity of $370 \pm 14$\kms and using the angular size of the nova found on Fig.~\ref{1}a, the distance is $d=525\pm28$ pc. The expansion velocity of the nova in the polar direction can also be estimated by converting the angular semi-major axis of $12.66\pm0.22$ arcsec to a physical size of $(9.94\pm0.63) \times 10^{14}$ m at this distance and then dividing by $t$. This gives $v_{pol}=501\pm32$\kms. Using the same method, the expansion velocity for the pinched equatorial belt is found to be $v_{eq}=340\pm14$\kms. 

Slit position 3 does not intersect with the bright central binary system and hence equatorial features in the spectra are more clearly visible. Fig.~\ref{2}b reveals variations in relative brightness between the \HA and [N{\sc ii}]6583\AA ~emission. By comparing emission intensities through aperture photometry, the \HA$\!\!$/[N{\sc ii}] ratio for the two circled features is 3.4, whilst the other clumps typically have a ratio of 9.1 with a standard deviation of 0.5. This is evidence for the presence of an equatorial ring enhanced in [N{\sc ii}] in DQ Her, as seen in FH Ser \citep{gill00}. \citet{petitjean90} also claim to have detected an [N{\sc ii}] equatorial ring around DQ Her. An explanation for the origin of these equatorial [N{\sc ii}] enhancements is yet to come (see \citealt{gill00}).

In addition, we note that in Fig.~\ref{2}b the blueshifted component of the equatorial ring lies below the faint stellar continuum while the redshifted component is situated above it. This can be interpreted as having a slit passing across a circular ring seen almost edge-on but tilted at a slight angle, which enables a measurement of the inclination of the shell. By measuring the vertical separation of the front and back components of the ring, the inclination is estimated at $86.8 \pm 0.2$ degrees, consistent with \citet{herbig92}. This is also in reasonable agreement with \citet{wood00}, who obtain their estimate of the inclination of the central binary system at $89.6\pm0.1$ degrees via smoothed particle hydrodynamic simulations. This is strong evidence for a direct relationship between the central binary and the structure of the extended remnant shell. Further evidence for this inclination can be found in \citet{slavin95} who measure the inclination of the shell to be $81\pm4$degrees from narrowband images of DQ Her showing different structural components of DQ Her. This implies, along with our results, that the inclination angle of the shell is somewhat smaller than that suggested by \citet{wood00}. Finally, the equatorial ring is most clearly visible on Fig. 1e of \citet{slavin95} which is taken through a filter whose bandpass is tuned to the emission from [N{\sc ii}] 6583 \AA. This is strong evidence for the existence of an [N{\sc ii}]-enhanced equatorial ring.

From the geometrical parameters described previously, the volume of the shell is $V = (2.24 \pm 0.14) \times 10^{45} ~\mathrm{m^{3}}$. Assuming that DQ Her has not been previously contaminated by other mass-loss processes and taking an average interstellar medium (ISM) density of $\rho = 10^{-21}$kg~m$^{-3}$, the corresponding swept-up mass is $\mathrm{M_{swept}} =(1.10\pm0.08) \times 10^{-6} \msun$. The mass of the ejected shell is believed to be $\mathrm{M_{shell}}=2.2 \times 10^{-4}\msun$ \citep{martin89}, much larger than the swept-up mass, and hence DQ Her should still be in the free expansion phase.

\begin{figure}
\begin{center}
\includegraphics[scale=0.5]{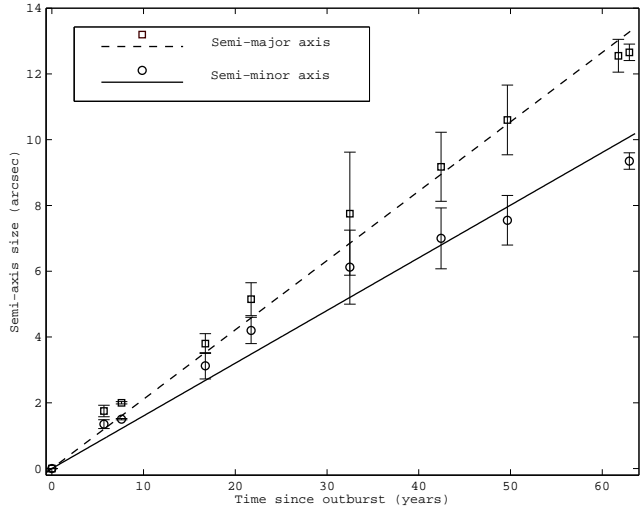}
\end{center}
\caption{The expansion of the shell of DQ Her as a function of time. Here we have added our measurements (last two points on the major axis curve and the last point on the minor axis curve) to figures found in the literature (\citealt{mustel70}, \citealt{williams78} and \citealt{duerbeck87}). As indicated, the squares and the dashed straight line fit represent the semi-major axis expansion whilst the circles and the solid straight line fit represent the semi-minor axis evolution.}
\label{4}
\end{figure}

Fig.~\ref{4} shows the expansion of the shell's semi-major and semi-minor axes as a function of time. The slopes of the straight line fits suggest expansion of the shell with constant velocities of $v_{major}=525\pm37$\kms ~for the poles and $v_{minor}=399\pm28$\kms ~for the equator. A likely reason why these velocities are slightly higher than our measurements made from the spectra is that the early size measurements (between $t = 0$ and $t = 40$ years) were made by measuring the outer angular radius of the shell, whereas in this work the midpoint of the emission was used to determine the size of the shell.

We compare the evolution of DQ Her to that of nova GK Per (1901), for which shell radii were also measured over time (see \citealt{anupama05}). The expansion velocity of GK Per remained approximately constant at $\sim1100\kms$ during the first 60 years. The mass contained in its ejected shell is believed to be of the order of $10^{-4}\msun$ (\citealt{balman99}; \citealt{balman05}), which is comparable to that of DQ Her. We can thus conclude that DQ Her, having a much smaller expansion velocity, should be in a free expansion phase for at least 60 years.

Studies of the nova shells of V603 Aql, GK Per, V476 Cyg and DQ Her \citep{duerbeck87} have suggested that they experience strong deceleration from their interaction with the ISM, with more rapidly expanding remnants slowing down the fastest. However three out of the four distances listed have since been revised (see \citealt{downes00} for GK Per and V476 Cyg, and this paper for DQ Her), having a direct impact on the velocity calculations. We also note that the half-lifetime (time after which the expansion velocity has dropped to half its initial value) quoted cannot be valid for GK Per and DQ Her. Indeed, from \citet{anupama05}, the initial expansion rate for GK Per is measured at $\sim1150\kms$, and after about 60 years (approximately the half-lifetime quoted), it is still $\sim 1080\kms$, a long way from half its initial value. Similarly in DQ Her where an initial expansion velocity of 325\kms ~is quoted, we find 370\kms ~today. Deceleration of expanding nova shells is expected from interaction with the ISM but our results along with \citet{anupama05} show that it tends to be much less significant than previously anticipated.

\begin{figure}
\begin{center}
\includegraphics[scale=0.6]{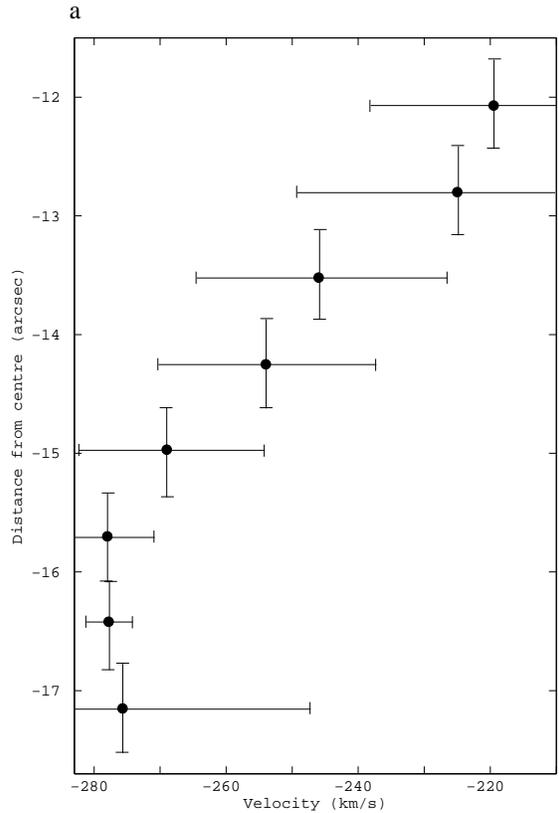}
\includegraphics[scale=0.6]{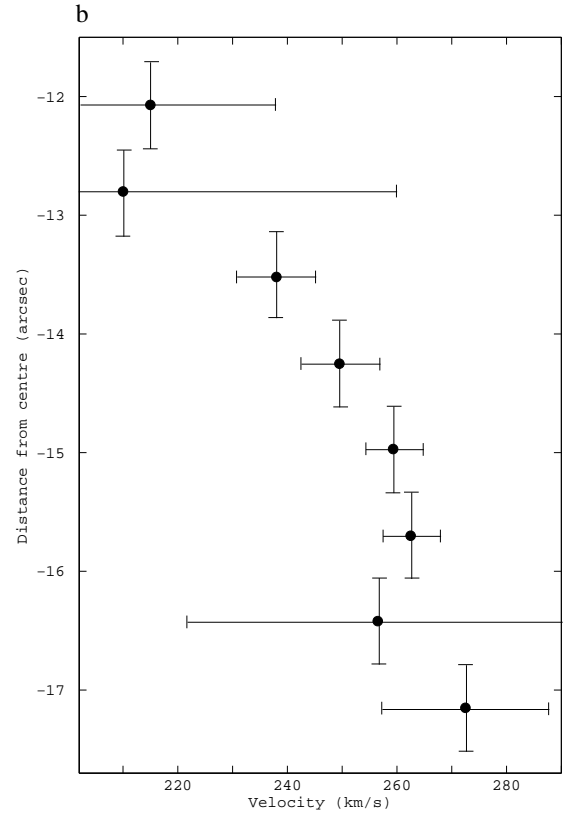}
\end{center}
\caption{Position-velocity arrays for tails 1 (a) and 3 (b). Line of sight velocities (x-axis), which have not been corrected for systemic velocity, are plotted against distance from the central star (y-axis). Both tails appear to reach a terminal velocity which is believed to be close to the projected wind velocity.}
\label{5}
\end{figure}

\subsection{Tails}\label{s4}

\begin{figure}
\begin{center}
\includegraphics[scale=.25]{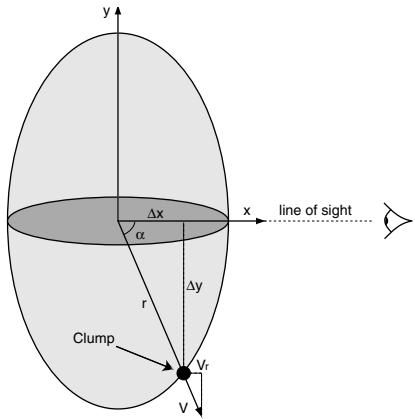}
\end{center}
\caption{Velocity projection along the line of sight for clumps on the surface of an ellipsoid. The $\Delta y$ position of the clumps was measured on the spectrum. The geometrical ellipse equation of the shell (found from Fig.~\ref{1}) was then used to calculate the angle $\alpha$. Because the spectra and image were taken only a year apart, the same ellipticity was used proportionally between the datasets which was a suitable approximation.}
\label{6}
\end{figure}

\citet{slavin95} were able to remove from their images the emission components that presented elliptical symmetry in order to discover extended tails. They argued that these filaments may originate either from an inner fast wind breaking through a clumpy shell or, alternatively, may represent regions which are shadowed by the dense clumps from the central photoionising radiation field. These tails are also clearly visible on Fig.~\ref{1}b, \ref{2} and \ref{3} where it is also evident that their velocities increase with distance from the centre. In the case of tail 1 (see Fig.~\ref{1}b), we can see that the magnitude of the line of sight velocity of the tail increases along its own length. This supports a model in which they are formed by wind ablation of clumps by a strong stellar wind. The top central clump in the \HA spectrum of position 2 (Fig.~\ref{2}a) also exhibits a structure resembling a bow-shock, which would be expected in the case of a wind flowing past clumps. Velocity profiles for these tails were extracted in order to determine whether the material is accelerating linearly or approaching a terminal velocity, which again would be anticipated in a situation where material is ablated from the clumps and joins the wind stream flowing past.

The results obtained for tails 1 and 3 are presented in Fig.~\ref{5}. This shows that the tails tend to attain a terminal velocity at their tips. The ratio of terminal velocity to wind stream velocity depends on the temperature contrast between the ablated and stream gas \citep{dyson06}, and it is thus not possible here to measure the wind velocity due to the lack of temperature data.  The terminal radial velocities of tails 1 and 3 were found to be $v_{1} = -280\pm15$\kms ~and $v_{3}=+265\pm15$\kms, respectively. The tails appear to be most clearly visible on the polar ends of the shell and not around the equator. This would also suggest the presence of a stellar wind that is collimated in the polar direction. Evidence for the presence of a wind in DQ Her has been discovered by \citet{eracleous98} through the study of the C{\sc iv} emission line and \citet{mukai03} from the study of scattered X-ray emission. \citet{selvelli03} also report the presence of a bipolar wind in nova HR Del (see however \citealt{slavin95} whose decomposed image of DQ Her (Fig. 2c) reveals equatorial tail features).

Absolute velocities for the tails were derived according to the following method. Our assumed geometry is an ellipsoidal shell with its minor axis along the line of sight and its major axis perpendicular to the line of sight (see Fig.~\ref{6}), which is the case for DQ Her since $i \approx 90^{\circ}$. A clump is situated on the surface of the shell lies at an angle $\alpha$ from the line of sight. A radially expanding tail emanating from this clump will have absolute velocity $v$ and line of sight velocity $v_{r}= v \cos \alpha$. By de-projecting the velocities measured on Fig.~\ref{5} ($\alpha = 72^{\circ} \pm 3^{\circ}$ for tails 1 and 3), the absolute terminal velocities of the tails are thus found to be $\sim 800-900$\kms, which could be close to the wind velocity.

\section{Discussion and conclusions}\label{s5}

High resolution longslit spectroscopy was performed on the shell of nova DQ Her, essential to accurately determine the expansion velocity of the remnant. The physical size of the ellipsoidal shell was then measured by inferring a constant equatorial expansion velocity, thus providing an accurate distance through expansion parallax. The size of the shell as a function of time is consistent with a constant expansion rate. The presence of accelerating tails emanating from the clumps around the poles of the shell suggests the existence of a fast ablating stellar wind collimated in the polar direction, rather than photoionisation-shadowed regions. Shock interaction between the clumps and the fast wind could also be an explanation for the localised extended X-ray emission regions observed by \citet{mukai03}.

We have considered possible evidence for abnormally high velocities of clumps close to the poles of the shell compared to that expected for a Hubble expansion. Given their current positions and velocities, they could have gone through an early slower velocity phase and were subsequently accelerated, possibly by the fast collimated wind. This could provide an attractive alternative mechanism for nova shell shaping (see \citealt{lloyd97}; \citealt{porter98} who consider the effects of a rotating WD envelope in the creation of prolate ellipsoidal shells). However, the high uncertainties on the velocities and positions of the clumps (mostly due to projection effects) along with the assumption of axisymmetry prevent us from presenting these results as firm conclusions.

Some aspects of the shell are still to be explained, such as the pinch observed at the equator and the [N{\sc ii}]-enhanced equatorial ring. As opposed to the tails discussed above, emission extending inside the shell is also visible on Fig.~\ref{3}. These inner clumps can clearly be seen in the spectra (Figs.~\ref{1} and \ref{2}), the most prominent example being the top clump on Fig.~\ref{2}b. Due to their position in the spectra, these clumps are either situated on the surface of the shell but have an abnormally low velocity, or are actually inside the shell, which is far more likely. The origin of such inner clumps and whether they are only located close to the poles or filling the whole shell volume is not clear.

\section*{Acknowledgements}

NMHV is supported by a University of Manchester research studentship and APR by a PPARC research studentship.
We would like to thank Deborah Mitchell for help with the data reduction and sensible comments on the writing of this paper, and Chris Gill for his preliminary work on DQ Her in his PhD thesis \citep{gill99}. WHT image of DQ Her (Fig.~\ref{1}a) taken by Vik Dhillon, University of Sheffield.

\label{lastpage}

\end{document}